\documentclass[%
 reprint,
 amsmath,amssymb,
 aps,
floatfix
]{revtex4-1}

\usepackage{scalerel,stackengine}
\usepackage{mathtools}
\usepackage[utf8]{inputenc}
\DeclarePairedDelimiter{\ceil}{\lceil}{\rceil}
\usepackage{graphicx}
\usepackage{dcolumn}
\usepackage{bm}
\usepackage{color}
\usepackage{xcolor,colortbl}
\usepackage[colorlinks,allcolors=blue]{hyperref}
\usepackage{booktabs}
\definecolor{Gray}{gray}{0.95}
\definecolor{LightCyan}{rgb}{0.95,1,1}
\stackMath
\newcommand\reallywidehat[1]{%
\savestack{\tmpbox}{\stretchto{%
  \scaleto{%
    \scalerel*[\widthof{\ensuremath{#1}}]{\kern+1.5pt\bigwedge\kern+1.5pt}%
    {\rule[-\textheight/4]{1ex}{\textheight}}%WIDTH-LIMITED BIG WEDGE
  }{\textheight}% 
}{0.5ex}}%
\stackon[1.5pt]{#1}{\tmpbox}%
}
\newcolumntype{a}{>{\columncolor{LightCyan}}c}
\newcolumntype{b}{>{\columncolor{white}}c}
\begin{document}
\preprint{APS/123-QED}

\title{Rapid Bayesian Inference of Global Network Statistics Using Random Walks}
%\thanks{A footnote to the article title}

\author{Willow B. Kion-Crosby}
\author{Alexandre V. Morozov}
%\email{Second.Author@institution.edu}
\affiliation{Department of Physics \& Astronomy, Rutgers, The State University of New Jersey.}

%\collaboration{MUSO Collaboration}%\noaffiliation

\date{\today}% It is always \today, today,

\begin{abstract}
We propose a novel Bayesian methodology which uses random walks for rapid inference of statistical properties of undirected networks with weighted or unweighted edges.
%The statistics of interest include, but are not limited to, the node degree distribution, the average degree of nearest-neighbor nodes, and the node clustering coefficient.
Our formalism yields high-accuracy estimates of the probability distribution of any network node-based property, and of the network size, after only a small fraction of network nodes has been explored.
The Bayesian nature of our approach provides rigorous estimates of all parameter uncertainties. We demonstrate our framework on several standard examples, including random, scale-free, and small-world networks, and apply it to study epidemic spreading on a scale-free network. We also infer properties of the large-scale network formed by hyperlinks between Wikipedia pages.
\end{abstract}

%\pacs{Valid PACS appear here}% PACS, the Physics and Astronomy
                             % Classification Scheme.
%\keywords{Suggested keywords}%Use showkeys class option if keyword
                              %display desired
\maketitle

%\tableofcontents

%\section{\label{sec:level1}Introduction}
Over the past few years, our lives have become increasingly dependent on large-scale networks. %, often available through our computers and smartphones.
In addition to the original computer-based networks such as the World Wide Web and the Internet, many online social networks have emerged, notably Twitter and Facebook.
Our professional and personal activities are influenced daily by knowledge-sharing online services such as Wikipedia and YouTube. More generally,
complex networks describe a broad spectrum of systems in nature, science, technology, and society~\cite{Albert2002,Newman2010}. Many of these networks are large and evolving, making
investigation of their statistical properties a challenging task. In particular, estimating the network size becomes non-trivial if the network is too large to visit every node. Consequently, predicting various network statistics, typically from random samples of limited size, has attracted considerable attention in the
literature~\cite{Newman2003,Lee2006,Yoon2007,Estrada2010,Gjoka2010,Cooper2012,Bliss2014,zhang2015}.\hfill

Here we develop a Bayesian approach to network sampling by random walks (RWs)~\cite{Yoon2007,Cooper2012}. Unlike previous results, our framework can be used to build
posterior probability distributions for any network node-based quantity of interest. Our approach reproduces several previously known global network statistics estimators within a single formalism,
automatically removes statistical biases caused by RW sampling~\cite{Yoon2007,Estrada2010}, and yields standard results in the uniform sampling limit.
Surprisingly, accurate estimates of various network properties, including its size, are obtained
after examining only a small fraction of all network nodes.

%The study of complex networks has been a topic of interest over recent years. The characterization of these networks is often done through the analysis of the network-wide averages of various statistics, e.g. node degrees, degree-degree correlations, clustering coefficients, etc.\cite{Bliss2014, Estrada2010, Yoon2007, Gjoka2010, Newman2003, Lee2006}. For these complex systems, querying entire networks can be cumbersome making the inference of global statistics from network samples preferable \cite{zhang2015, Cooper2012}. In this work we have developed a Bayesian formalism based on the network sampling method of random walks (RWs). With this methodology and theoretical framework we can obtain not only the network-wide averages of these statistics, but also their full distributions for systems in which only the walker's local environment is known. Our framework naturally removes biases previously known to be introduced by the sampling method of RWs \cite{Estrada2010, Yoon2007}, which is often done by-hand, and gives rise to a rigorous network property estimation error analysis in the form of full posterior probability distributions. Additionally, this framework reproduces several previously-known global network statistics estimators within a single formalism. We have demonstrated the fidelity of this formalism on various model systems where the true network properties can be compared, as well as on the network formed by hyperlinks between pages on Wikipedia. \hfill

Consider a RW on a network of $N$ nodes with weighted edges: $\{w_{ji}\}$, where $w_{ji}$ is the rate of transition from node $i$ to node $j$.
At each step the walker will transition to a neighboring node with a probability $P(i \to j) = w_{ji} / \sum_{k \in \{nn\}_i} w_{ki}$, where the sum is over all nearest neighbors of node $i$.
We subdivide all network nodes into sets $S_x$ based on the value of some property $x$, such as the number of links connected to the current node, known as the node degree~\cite{Albert2002}; there are $N_x$ nodes in each set and ${\cal N}_x$ distinct sets. We assume that the property in question is discrete; continuous properties can be discretized by binning.
We focus on undirected networks with symmetric rates, $w_{ji} = w_{ij}$.
In this case, the stationary probability for the RW to occupy node $i$, $\pi_i$, can be determined using the steady-state master equation~\cite{vanKampen2007,Krapivsky2010}:
\begin{equation} \label{eq:master}
\sum_{j \in \{nn\}_i} [\pi_j P(j \to i) - \pi_i P(i \to j)] = 0.
\end{equation}
Equation~\eqref{eq:master} is satisfied if $\pi_i \sim w_i = \sum_{k \in \{nn\}_i} w_{ki}$, where $w_i$ is the total outward rate.
For unweighted networks, the node's stationary probability is proportional to its degree $k_i$~\cite{Noh2004}.
With normalization, the stationary probabilities become $\pi_i = w_i/\sum_{i=1}^N w_i$.

If the walker starts from a node with property $x$, the average number of steps between subsequent visits to any node within the set $S_x$,
also known as the mean return time (RT), is given by~\cite{Condamin2007}:
\begin{equation} \label{eq:return_time}
\langle \ell \rangle_x = \frac{1}{\sum_{i\in S_x }\pi_i}.
\end{equation}
In the case of undirected networks,
\begin{equation}\label{eq:MRT}
\langle \ell \rangle_x = \frac{\langle w \rangle}{p_x \langle w \rangle_x},
\end{equation}
where $p_x = N_x / N$ is the fraction of nodes with property $x$, $\langle w \rangle = N^{-1} \sum_{i=1}^N w_i$, and
$\langle w \rangle_x = N_x^{-1} \sum_{i=1}^{N_x} w_i$.

The probability of making $\Delta \ell$ steps between subsequent visits to $S_x$, $P(\Delta \ell)$, is asymptotically exponential in arbitrary finite networks~\cite{Bollt2005}:
\begin{equation} \label{exp_dist}
%P(\Delta \ell) \simeq q_x (1 - q_x)^{\Delta \ell - 1} \simeq q_x e^{- q_x \Delta \ell} \; \text{ for } \; \Delta \ell = 1,2, \dots,
P(\Delta \ell) \simeq q_x e^{- q_x \Delta \ell} \; \text{ for } \; \Delta \ell = 0,1, \dots,
\end{equation}
where $q_x = \langle \ell \rangle_x^{-1} \ll 1$ is the hitting rate of the nodes within $S_x$.
We find empirically that the exponential ansatz for $P(\Delta \ell)$ is sufficiently accurate for our purposes, % (Fig.~\ref{fig:TDE_fig}(a)--\ref{fig:wikipedia_fig}(a)),
although in principle our approach is not limited to it.
The likelihood that during a single RW with $\ell \gg 1$ steps the walker has visited $\mathcal{K}_x$ nodes in $S_x$
is then given by the Poisson distribution:
\begin{equation} \label{eq:likelihood}
P(\mathcal{K}_x|q_x) = \frac{(q_x \ell)^{\mathcal{K}_x} }{\mathcal{K}_x!} e^{-q_x\ell}.
\end{equation}
This likelihood function is maximized by $\hat{q}_x = \mathcal{K}_x / \ell$, which implies $\mathcal{K}_x \ll \ell$.
Assuming a uniform prior for $q_x$ in the $[0,1]$ range, the posterior probability density for $q_x$ becomes
\begin{equation} \label{eq:posterior}
P(q_x| \mathcal{K}_x) = \frac{1}{B(\mathcal{K}_x,\ell)} q_x^{\mathcal{K}_x} e^{-q_x \ell},
\end{equation}
where $B(\mathcal{K}_x,\ell) = \int_0^1 dq_x q_x^{\mathcal{K}_x} e^{-q_x \ell} \simeq \mathcal{K}_{x}!/\ell^{\mathcal{K}_x+1}$ is a normalization constant.
%and the approximation is valid in the $\ell \gg 1$ limit.
Thus Eq.~\eqref{eq:posterior} is closely approximated by a gamma distribution $\Gamma(q_x;\mathcal{K}_x + 1, \ell)$, which becomes Gaussian in the $\mathcal{K}_x \gg 1$ limit,
with the mean $\bar{q}_x = \hat{q}_x$ and the standard deviation $\sigma_{q_x} = \hat{q}_x / \sqrt{\mathcal{K}_x}$.

This result in combination with Eq.~\eqref{eq:MRT} yields a maximum likelihood estimate (MLE) and a standard error for the probability $p_x$ of the property $x$:
\begin{equation} \label{eq:ML_distr}
%\hat{p}_x = \frac{\langle w \rangle \mathcal{K}_x}{\langle w \rangle_x\ell}\; \text{ and }\; \sigma_{p_x} = \frac{\hat{p}_x}{\sqrt{\mathcal{K}_x}}.
\hat{p}_x = \frac{\mathcal{K}_x/\langle w \rangle_x}{\sum_x^{{\cal N}_x} \mathcal{K}_x /\langle w \rangle_x}\; \text{ and }\; \sigma_{p_x} = \frac{\hat{p}_x}{\sqrt{\mathcal{K}_x}}.
\end{equation}
If the property of the node $i$ is its total outward rate $w_i$ discretized into ${\cal N}_w$ bins, Eq.~\eqref{eq:ML_distr} yields
%imposing the normalization requirement on $\hat{p}_w$, $\sum_{j=1}^{{\cal N}_w} \hat{p}_{w_j} = 1$, yields an estimate for the total outward edge weights averaged over the entire network:
\begin{equation} \label{eq:p.w}
%\frac{\langle w \rangle}{\ell} =  \left( \sum_{j=1}^{{\cal N}_w} \frac{\mathcal{K}_{w_j}}{w_j} \right)^{-1}.
\hat{p}_{w_i} = \frac{\mathcal{K}_{w_i} / w_i}{\sum_{j=1}^{\mathcal{N}_w} \mathcal{K}_{w_j} / w_j},
\end{equation}
where $\mathcal{K}_{w_j}$ is the number of visits to nodes with total outward rates in the bin $j$.
For unweighted networks ($w_{ij}=1, \; \forall i,j$), $\hat{p}_{w_i}$ reduces to $\hat{p}_{k_i}$, the network degree distribution~\cite{Albert2002}.

For an arbitrary node property $x$, each set $S_x$ can be additionally subdivided by the binned value of $w$, such that
\begin{equation} \label{eq:net_dists}
\hat{p}_x = \sum_{j=1}^{{\cal N}_w}  \hat{p}_{x,w_j} =   %\frac{\langle w \rangle}{\ell}  \sum_{j=1}^{{\cal N}_w}  \frac{\mathcal{K}_{x,w_j} }{w_j} =
\sum_{j=1}^{{\cal N}_w} \frac{\mathcal{K}_{x,w_j} }{w_j} / \sum_{j=1}^{{\cal N}_w} \frac{\mathcal{K}_{w_j}}{w_j}.
\end{equation}
%where Eq.~\eqref{eq:p.w} was employed to compute $\hat{p}_{x,w_j}$.
Here, $\mathcal{K}_{x,w_j}$ is the number of visits to nodes with both property $x$ and the total outward rates in the bin $j$. Thus, the knowledge of
$\mathcal{K}_{w_j}$,  $\mathcal{K}_{x,w_j}$, and $w_j$ is sufficient to compute
the MLE of any property $x$ and estimate its uncertainty (Eq.~\eqref{eq:ML_distr}).
Note that the division by $w_j$ in Eq.~\eqref{eq:net_dists} corrects for the bias introduced by RW sampling~\cite{Estrada2010, Yoon2007, Gjoka2010}.
%If $x = w_i$, $\mathcal{K}_{w_i,w_j} = \mathcal{K}_{w_i} \delta_{ij}$, in agreement with Eq.~\eqref{eq:p.w}.

The MLE of the average outward rate is given by
\begin{equation}\label{eq:rate_est}
\hat{\langle w \rangle} = \sum_{i=1}^{\mathcal{N}_w} w_i \hat{p}_{w_i} = \frac{\ell}{\sum_{j=1}^{\mathcal{N}_w} \mathcal{K}_{w_j} / w_j}, % \; \text{ and }\; \sigma_{\langle \mu \rangle} = \sqrt{\ell} \big / \sum_{\mu'} \frac{\mathcal{K}_{\mu'}}{\mu'}.
\end{equation}
where we used $\sum_{i=1}^{\mathcal{N}_w} \mathcal{K}_{w_i} = \ell$.
The uncertainty of this estimate can be evaluated using $\sigma_{\langle w \rangle}^2 = \sum_{i=1}^{\mathcal{N}_w} w_i^2 \sigma^2_{p_{w_i}}$ as well as Eqs.~\eqref{eq:ML_distr} and \eqref{eq:p.w}, to yield $\sigma_{\langle w \rangle} = \hat{\langle w \rangle}/\sqrt{\ell}$, in accordance with the central limit theorem. Similarly, for an arbitrary property $x$
\begin{equation}
\hat{\langle x \rangle} = \sum_x x \hat{p}_x \; \text{ and } \; \sigma^2_{\langle x \rangle} = \sum_{x} x^2 \sigma_{p_{x}}^2. 
\end{equation}
%\subsection{Network size estimate}

Let us now suppose that the network nodes are divided into two sets: $N_p$ randomly chosen nodes, which we shall refer to as \emph{pseudotargets}, and all the rest.
The pseudotarget nodes are drawn prior to exploring the network, so that their average outward rate, $\langle w \rangle_p$, is known. Equations \eqref{eq:MRT} and \eqref{eq:likelihood} can now be used to construct the posterior probability for the network size
(assuming a uniform prior in the $[N_p,N_{max}]$ range, where $N_{max}$ denotes an upper limit on $N$):
\begin{equation} \label{eq:N_post}
P(N|\mathcal{K}_p) = \frac{N^{-\mathcal{K}_p} \exp \left\{ -\frac{N_p \langle w \rangle_p}{N \langle w \rangle} \ell \right\} }{\sum_{\tilde{N} = N_p}^{N_{max} } \tilde{N}^{-\mathcal{K}_p} \exp \left\{ -\frac{N_p \langle w \rangle_p}{\tilde{N} \langle w \rangle} \ell \right\} },
%=\left[\sum_{N' = N_p}^{N_{max}}\left(\frac{N}{N'}\right)^{\mathcal{K}_p} \exp\left\{-\frac{N_p \langle \mu \rangle_p \ell(N - N')}{ \langle \mu \rangle N N'}\right\}\right]^{-1},
\end{equation}
where $\mathcal{K}_p$ is the number of visits to pseudotargets.
Note that using uniform priors in both Eqs.~\eqref{eq:posterior} and~\eqref{eq:N_post} does not affect
the results as long as $\mathcal{K}_x$ and $\mathcal{K}_p$, respectively, are sufficiently large.  
Similar to Eq.~\eqref{eq:posterior}, we find that this posterior probability rapidly becomes Gaussian as $\mathcal{K}_p$ increases, with
%The saddle-point approximation then yields
\begin{equation}\label{eq:N_ML}
\hat{N} = \frac{\ell N_p \langle w \rangle_p}{\mathcal{K}_p \langle w \rangle}\;\text{ and }\; \sigma_{N} = \frac{\hat{N}}{\sqrt{\mathcal{K}_p}}.
\end{equation}
Using Eq.~\eqref{eq:rate_est}, we obtain
\begin{equation} \label{eq:network_size}
\hat{N} = \frac{N_p \langle w \rangle_p}{\mathcal{K}_p} \sum_{j=1}^{\mathcal{N}_w} \frac{\mathcal{K}_{w_j} }{w_j}.
\end{equation}
Note that the error in $\hat{N}$ can be reduced either through increasing $N_p$ or assigning highly-connected nodes (network hubs) to be pseudotargets.
In the $N_p = 1$ case, Eq.~\eqref{eq:N_ML} recovers the network size estimator from Ref.~\cite{Cooper2012}.
%% Discuss this in Intro: %%
%This process of counting returns to pseudotargets is similar to the methods for computing network size estimators discussed in both \cite{Katzir2015} and \cite{Cooper2012}. \hfill

Note that in the case of a complete unweighted network in which each node is connected to all $N$ nodes (including itself),
RW sampling reduces to uniform sampling with replacement. In this limit, Eq.~\eqref{eq:ML_distr} yields $\hat{p}_x = \mathcal{K}_x / \ell$ and $\sigma_{p_x} = \sqrt{\mathcal{K}_x} / \ell$,
consistent with the standard results based on binomial sampling. Moreover, $\hat{N} = {\ell N_p}/{\mathcal{K}_p}$ in this case, reproducing the classic Lincoln-Petersen estimator of
biological population sizes by the mark and recapture method~\cite{Webster2013} (the differences between uniform sampling with and without replacement can be neglected in the $\mathcal{K}_p \ll N_p$
limit). These results remain valid for any network in which the total outward rate $w$ is the same for every node. Note that the key difference between RW sampling and uniform sampling is that
the former preferentially visits the nodes with larger $w$ values, so that, given $\ell$, $\sigma_{p_x}$ is smaller for RW if $\langle w \rangle_x > \langle w \rangle$, and vice versa.

We have implemented the above theoretical framework as follows: for each network, $N_p$ pseudotargets are randomly drawn and their $\langle w \rangle_p$ is computed. Commencing the RW from one of these pseudotargets, we record $\ell$, $\mathcal{K}_p$, $\{\mathcal{K}_{w}\}$, and $\{\mathcal{K}_{x,w}\}$ for a desired set of node properties $x$. At each step in the RW, Eqs.~\eqref{eq:ML_distr}--\eqref{eq:network_size} can then be used to infer various network properties.
\setlength{\arrayrulewidth}{.8pt}
\begin{table*}[t]
  \centering
  \caption{TDE model statistics summary. Shown are MLE and $95\%$ confidence interval ($\pm 2 \sigma$) for each quantity, followed by exact values for the TDE model system. All predictions are based on a single representative RW with $\ell = 10^4$ steps corresponding to the unit time interval in the TDE model.} \label{tab:1}
\begin{tabular}{cccccccccc}
%{p{2cm}p{1.8cm}p{.6cm}p{1.3cm}p{.8cm}p{1.6cm}p{1.3cm}p{2cm}p{2cm}p{2cm}p{1.6cm}  }
%
 \hline
 \hline
   %\rowcolor{Gray}
   & & & & & & & & & \\
   %\rowcolor{Gray}
   $\hat{N}$ & $N$ & $\hat{\langle k \rangle}$ & $\langle k \rangle$ & $\reallywidehat{\langle \langle k_{nn} \rangle \rangle}$  & $\langle \langle k_{nn} \rangle \rangle$ & $\hat{\langle C \rangle}$ & $\langle C \rangle$ &  $\reallywidehat{W/N} $ & $W/N$ \\
   %\rowcolor{Gray}
  $\pm 2 \sigma_N$ & & $\pm 2 \sigma_{\langle k \rangle}$ & & $\pm 2 \sigma_{\langle\langle k_{nn} \rangle \rangle}$ & & $\pm 2 \sigma_{\langle C \rangle}$ & & $\pm 2 \sigma_{W/N}$ & \\ \\
 \hline
   & & & & & & & & & \\
  $1.01 \times 10^5$   & $10^5$ & $8.02$ & $8.14$ & $64.6$ & $67.1$ & $0.251$ & $0.255$ & $2.000$ & $2.000$ \\
 
  $\pm$ $0.08 \times 10^5$ & & $\pm 0.16$ & & $ \pm4.2$ & & $ \pm0.011$ & & $\pm 0.045$ & \\
  & & & & & & & & & \\
 \hline
 \hline
\end{tabular}
\end{table*} %\hfill

To verify the validity of our algorithm on standard model systems, we have studied three unweighted, undirected networks: an Erd\H{o}s-R\'enyi (ER) random graph~\cite{Erdos1961}, a scale-free (SF) random graph~\cite{Albert2002}, and a small-world (SW) network~\cite{Watts1998}. Each network has $N = 10^6$ nodes. The ER network was constructed by randomly assigning $\ceil{N\log (N) / 2}$ edges between nodes, the SF network by the preferential attachment method~\cite{Albert2002} with $m = 2$ edges attached to new nodes, and the SW network as described in Ref.~\cite{Newman2000}, with the shortcut probability $p=1/2$.

For each network, $N_p = 10^3$ pseudotargets were randomly drawn and the network was subsequently explored with a RW for $\ell = 10^5$ steps,
visiting at most 10\% of all nodes.
Besides network size and the node degree distribution, we have focused on posterior probabilities of the average degree of nearest-neighbor nodes, which is a measure of network degree assortativity,
\begin{equation} \label{eq:knn}
\langle k_{nn} \rangle_i \equiv k_i^{-1} \sum_{j\in \{nn\}_i} {k_j},
\end{equation}
the clustering coefficient~\cite{Newman2003}, 
\begin{equation} \label{eq:Ci}
C_i \equiv \frac{2y}{k_i(k_i-1)},
\end{equation}
where $y$ is the total number of links shared by the nearest neighbors of node $i$, and a measure of the degree inhomogeneity~\cite{Estrada2010} 
\begin{equation} \label{eq:rhoi}
\rho_i \equiv \sum_{j \in \{nn\}_i} \left(k_i^{-1/2} - k_j^{-1/2}\right)^2.
\end{equation}

A comprehensive summary of the inferred network statistics can be found in the Supplementary Material (SM)~\cite{SM2018}. 
Although network topologies of these three systems are quite different, all statistics we have considered are predicted accurately.
%A summary of the ER system statistics is provided in Fig.~\ref{fig:erdosrenyi_fig}. Fig.~\ref{fig:erdosrenyi_fig}(a) shows that the exponential ansatz for the RT distribution,
%Eq.~\eqref{exp_dist}, is accurate for this system. Fig.~\ref{fig:erdosrenyi_fig}(b) demonstrates the convergence of the average degree distribution and the network size to the exact values during 5
%representative runs. The predicted degree distribution, $p_{k_i}$, known to be  Poisson~\cite{Erdos1961}, is shown in Fig.~\ref{fig:erdosrenyi_fig}(c). Finally, in Fig.~\ref{fig:erdosrenyi_fig}(d)
%we demonstrate evolution of the posterior distribution for the network size as more data is collected. Additional statistics for the ER, SF and SW systems are summarized in Table~\ref{tab:1}. Although the network topologies of these three systems are quite different, all network statistics we have considered are recovered accurately.

Next, we have constructed a generalized ER network with $N=10^6$ nodes and weighted edges. After placing all the edges as in the unweighted ER network, a loop was added to each node with probability $p = 1/2$. All loops and edges were then assigned a symmetric weight $w_{ij} = w_{ji}$ drawn from an exponential distribution with unit mean. For this system, we have collected statistics on each node's total outward rate, $w_i$, loop weight, $w^{\mathrm{loop}}_i = w_{ii}$ (note that $w_{ii} = 0$ for nodes without loops), outward rate averaged over all nearest neighbors of node $i$, $\langle w_{nn}\rangle_i$, and average nearest-neighbor loop weight, $\langle w_{nn}^{\mathrm{loop}} \rangle_i$.
We have again employed a RW with $\ell = 10^5$ steps and $N_p = 10^3$ randomly drawn pseudotargets.
Although the RT distribution for this system deviates from purely exponential due to loops,
all the network statistics we have considered are again predicted accurately~\cite{SM2018}. Thus our methodology is equally applicable to studies of weighted networks with loops.

%Next, we have constructed a generalized ER network with $N=10^6$ nodes and weighted edges. After placing all the edges as in the unweighted ER network, a loop was added to each node with probability $p = 1/2$. All loops and edges were then assigned a symmetric weight $w_{ij} = w_{ji}$ drawn from an exponential distribution with unit mean. For this system, we have collected statistics on each node's total outward rate, $w_i$, loop weight, $w^{\mathrm{loop}}_i = w_{ii}$ (note that $w_{ii} = 0$ for nodes without loops), outward rate averaged over all nearest neighbors of node $i$, $\langle w_{nn}\rangle_i$, and average nearest-neighbor loop weight, $\langle w_{nn}^{\mathrm{loop}} \rangle_i$.
%(\textcolor{red}{is this average only for nodes with loops? no, all nodes}).

%We have explored the statistics of these quantities using a RW with $\ell = 10^5$ steps and $N_p = 10^3$ randomly drawn pseudotargets (final row of Table~\ref{tab:1}, Fig.~\ref{fig:Gen_ER_fig}).
%Note that the RT distribution for this system deviates from purely exponential since many returns occur after a single step due to loops (Fig.~\ref{fig:Gen_ER_fig}(a)). Nonetheless, 
%all the network statistics we have considered are predicted accurately (Fig.~\ref{fig:Gen_ER_fig}(b)--(d)), except for the tail of the Fig.~\ref{fig:Gen_ER_fig}(d) distribution since those rare events were not observed.
%Thus our methodology is equally applicable to studies of weighted networks with loops.

After validating our approach on model systems, we have demonstrated its effectiveness in a more realistic setting, by tracking an epidemic spreading on a scale-free network
in the traffic-driven epidemiological (TDE) model~\cite{Meloni2009}. Following Ref.~\cite{Meloni2009}, we have generated the underlying network using a hidden-metric approach, which employs a tunable
parameter $\alpha$ to control the degree of local node clustering~\cite{Serrano2008}.
The number of links in each node is drawn from a power-law distribution, $p_{k_i} \sim {k_i}^{-\gamma}$. For our network, we have chosen $N = 10^5$, $\gamma = 2.6$, and $\alpha = 2$ (which leads to significant
clustering).

\begin{figure}[h]
\centering
\includegraphics[width=0.49\textwidth]{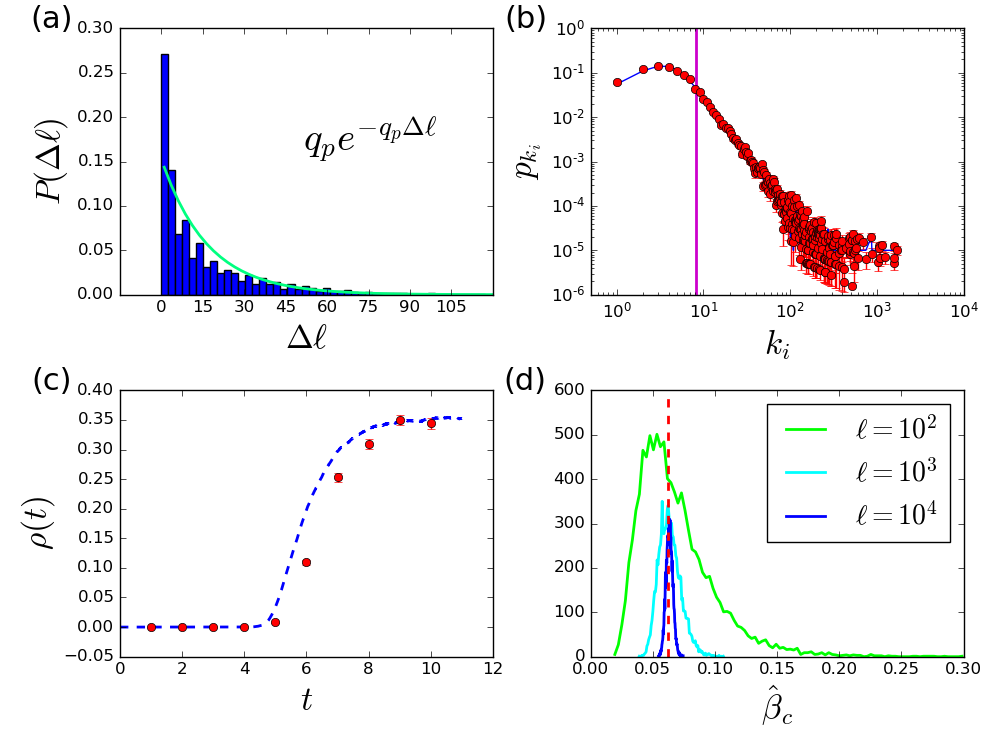}
\caption{Epidemic spreading statistics.
(a) Pseudotarget RT distribution. Equation~\eqref{exp_dist} parameterized by exact $q_p$ is shown in cyan.
(b) $\mathrm{MLE} \pm 2 \sigma$ (red circles with error bars) for the node degree distribution; exact distribution is shown in blue and its average is shown as a vertical line. 
(c) $\mathrm{MLE} \pm 2 \sigma$ (red circles with error bars) for the fraction of infected nodes $\rho(t)$ computed at unit time intervals, with the exact value shown as a dashed blue curve.
(d) Histograms of $\beta_c$ MLEs obtained using $10^4$ independent runs with $\ell = 10^2$, $10^3$, $10^4$ steps. Exact value is shown as a vertical dashed line.
} \label{fig:TDE_fig}
\end{figure}

Epidemic propagation was simulated through the exchange of $W$ contagion packets between nodes (see Ref.~~\cite{Meloni2009} for details).
Briefly, each node can be in either susceptible or infected state; the simulation starts with a single infected node.
When a packet moves from node $i$ to node $j$ on the network, node $j$ becomes infected with the spreading
probability $\beta$ if node $i$ was infected; infected nodes can also recover with rate $\mu$, set to $1$ without loss of generality. We have focused on the case in which contagion packets perform RWs between randomly assigned initial and destination nodes. Once a packet reaches its destination, it is removed and a new packet is added to keep $W$ constant.
On average, each packet moves once per unit simulation time. Under this choice of packet dynamics, there is a critical value of
$\beta_c = (\langle k \rangle^2 / \langle k^2 \rangle) N/W$ above which a sustained epidemic outbreak is observed~\cite{Meloni2009}. We have set $W=2N$ and $\beta = 7 \times 10^{-1} \gg \beta_c = 6.24 \times 10^{-2}$ in the simulation.

We have used a single RW with $\ell = 10^4$ steps and $N_p = 10^3$ pseudotargets to verify the validity of our exponential ansatz (Fig.~\ref{fig:TDE_fig}(a)) and predict the node degree distribution % $p_{k_i}$
(Fig.~\ref{fig:TDE_fig}(b)); several other statistics relevant to the study of epidemics on networks~\cite{Pellis2015} are listed in Table~\ref{tab:1}.
In addition, we have tracked time-dependent evolution of the fraction of infected nodes $\rho(t)$ (Fig.~\ref{fig:TDE_fig}(c)). We have assumed that nodes can be queried much faster than the
time scales on which the epidemic spreads, and thus matched $\ell$ steps of our RW sampling to the unit time interval in the TDE model (Fig.~\ref{fig:TDE_fig}(c), Table~\ref{tab:1}). Finally, we have predicted $\beta_c$ using the evolving system's snapshot,
again under the assumption that RW sampling is fast compared to the time scales of the epidemics (Fig.~\ref{fig:TDE_fig}(d)). % "frozen" assumption

%The MLE and $95\%$ confidence intervals for the network size, average degree, average nearest-neighbor degree, clustering coefficient, and average packet occupancy from a single representative sampling of $\ell = 10^4$ steps are displayed in Table~\ref{tab:1} with the full degree distribution shown in Fig.~\ref{fig:TDE_fig}(b). Additionally the estimation of $\rho(t)$ at the end of each of the intervals alongside the true simulation value are shown in Fig.~\ref{fig:TDE_fig}(c). Note that during the interval $t=4...8$ when the epidemic spread is the most rapid, the estimation is below the true value as the statistic is based on a range of values of $\rho(t)$.

%{\color{red} comment about immunization} \\
%{\color{red} comment about applying this process to network of cities or airlines or viruses spreading on the WWW} \\

\begin{figure}[h]
\centering
\includegraphics[width=0.49\textwidth]{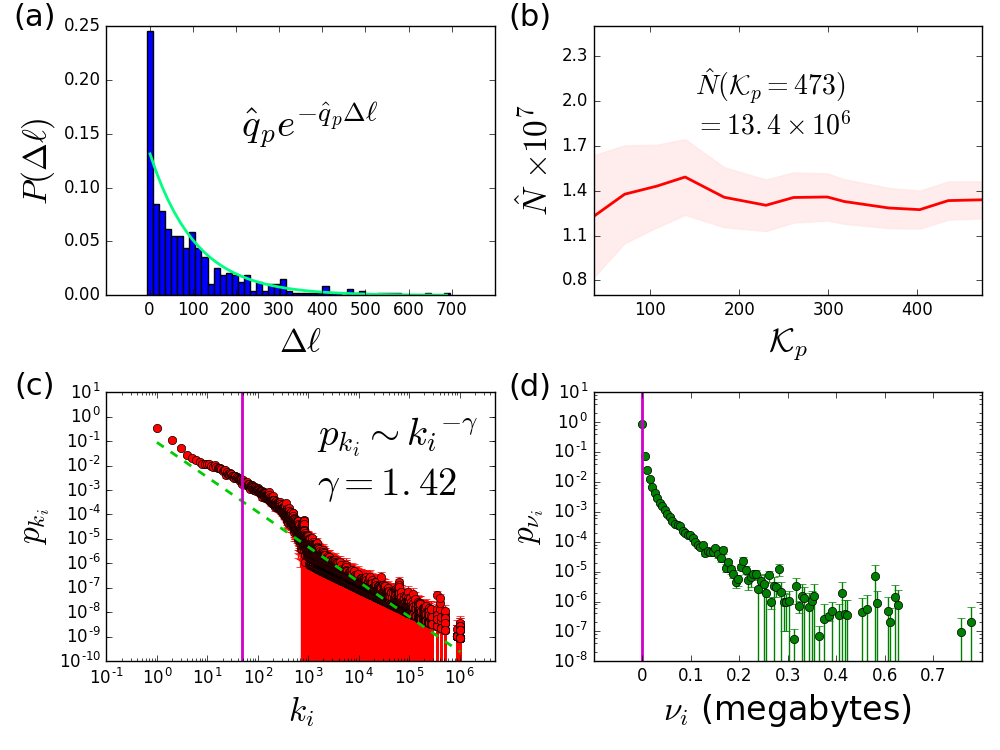}
\caption{
Wikipedia network statistics.
(a) Pseudotarget RT distribution. Equation~\eqref{exp_dist} parameterized by $\hat{q}_p$ is shown in cyan.
%Distribution of RTs between subsequent visits to pseudotargets. Exponential distribution parameterized by the MLE of the hitting rate $\hat{q}_p$ is plotted in cyan.
(b) $\mathrm{MLE} \pm 2 \sigma$ for $N$ as a function of ${\cal K}_p$.
%Evolution of the posterior probability for the network size. Shown are the MLE (red curve) and the 95\% confidence interval.
(c) $\mathrm{MLE} \pm 2 \sigma$ for the degree distribution of Wikipedia pages of all types. Power-law fit is shown as a green dashed line. % $p_{k_i} \sim {k_i}^{-\gamma}$
Average degree is shown as a vertical line.
(d) $\mathrm{MLE} \pm 2 \sigma$ for the distribution of Wikipedia page sizes. Average size is shown as a vertical line.
} \label{fig:wikipedia_fig}
\end{figure} %\hfill

\begin{table*}[t]
  \caption{Wikipedia statistics summary. Shown are MLE$\pm 2 \sigma$ for each quantity. All predictions are based on a single trial with $\ell = 5 \times 10^4$ steps.
  $N_a$ and $N_r$ are the total number of English articles and redirect pages in Wikipedia, as of Dec. 2017~\cite{WikiPage2}.}
  \label{tab:3}
\begin{tabular}{ccccccccccc}
%{p{2cm}p{1.8cm}p{.6cm}p{1.3cm}p{.8cm}p{1.6cm}p{1.3cm}p{2cm}p{2cm}p{2cm}p{1.6cm}  }
%
 \hline
 \hline
   %\rowcolor{Gray}
   & & & & & & & & & & \\
   %\rowcolor{Gray}
   $\hat{N}$ & $\hat{\langle k \rangle}$ & $\reallywidehat{\langle \chi_r \rangle}$ & $\reallywidehat{\langle \chi_d \rangle}$ & $\reallywidehat{\langle \nu_a \rangle}$ & $\hat{\langle \nu \rangle}$ & $\hat{N}(1 - \reallywidehat{\langle \chi_r \rangle})$ & $N_a$ & $\hat{N} \reallywidehat{\langle \chi_r \rangle}$ & $N_r$ & $\hat{N} \reallywidehat{\langle \nu_a \rangle}$ \\
   %\rowcolor{Gray}
  $\pm 2 \sigma_N$ & $\pm 2 \sigma_{\langle k \rangle}$ & $\pm 2 \sigma_{\langle \chi_r \rangle}$ & $\pm 2 \sigma_{\langle \chi_d \rangle}$ & $\pm 2 \sigma_{\langle\nu_a \rangle}$ & $\pm 2 \sigma_{\langle\nu \rangle}$ & $\pm 2 \sigma_{N (1 - \langle \chi_r \rangle)}$ & & $\pm 2 \sigma_{N \langle \chi_r \rangle}$ & & $\pm 2 \sigma_{N \langle \nu_a \rangle}$ \\ \\
 \hline
  & & & & & & & & & &  \\
 $13.4 \times 10^6$ & $47.7$ & $0.6009$ & $0.0399$ & $2670$ & $2720 $ & $5.35 \times 10^6$ & $5.3 \times 10^6$ & $8.05 \times 10^6$ & $8.0 \times 10^6$ & $35.8$ \\
 $\pm 1.2 \times 10^6 $ & $\pm 0.4$ & $\pm 0.0197$ & $\pm 0.0047$ & $\pm 40 \text{ bytes}$ & $\pm 40 \text{ bytes}$ & $\pm 0.56 \times 10^6$ & & $\pm 0.79 \times 10^6$ & & $\pm 3.3 \text{ GB}$ \\
  & & & & & & & & & &  \\
 \hline
 \hline
\end{tabular}
\end{table*} %\hfill

Next, we have examined the network formed by hyperlinks between English articles on Wikipedia. %using Python modules urllib and HTMLParser.
Links connecting an article to itself were disregarded, multiple links between articles were counted as one, and automatic redirects were disallowed, resulting in an unweighted, undirected, loopless network
consisting of all English articles, redirect pages, and disambiguation pages~\cite{WikiPage}.
%For a definition of what is considered an English article see \url{https://en.wikipedia.org/wiki/Wikipedia:What_is_an_article?}
To assign pseudotargets, the first $5000$ pages were drawn from Wikipedia's static HTML dumps. A single randomly chosen link was then taken from each of these pages and the node it pointed to was designated as a pseudotarget,
resulting in $N_p = 4769$. This procedure increases the likelihood that the pseudotargets are hubs with a large number of links, facilitating
collection of the network statistics since $\mathcal{K}_p$ grows more rapidly~\cite{Noh2004,Lee2006,Cooper2012}.
%Artificially increasing $\langle k \rangle_p$ in this way significantly augments the rate of $\mathcal{K}_p$ accumulation and so causes the error in the network size estimate to diminish rapidly. In theory further hubs could be sought to increase the rate at which $\sigma_N \rightarrow 0$, however, eventually this would lead to the exponential approximation of the RT distribution to breakdown. Even so, a pseudotarget set with $q_p$ close to $1$ would still yield an accurate network size estimate as is clear upon examination of Eq.~\eqref{eq:N_ML}.\hfill

We have focused on several statistics that facilitate comparison with known properties of Wikipedia: the size of each page in bytes, $\nu$ (as provided by Wikipedia), and two variables $\chi_r,\chi_d \in \{0,1\}$ representing whether a page is a redirect or a disambiguation page, respectively.  The quantities $\langle \chi_r \rangle$, $\langle \chi_d\rangle$, and $\langle \nu_a \rangle \equiv \langle (1-\chi_r) \nu \rangle$ then give the fraction of redirect pages, disambiguation pages,
and the average storage space in bytes of English articles (Wikipedia excludes redirect pages from
its estimates of the number of articles~\cite{WikiPage}). The RW was run for $\ell = 5 \times 10^4$ steps, with the resulting predictions shown in Table~\ref{tab:3} and Fig.~\ref{fig:wikipedia_fig}.

We find that Wikipedia contains $13.4$ million pages, each of which is connected on average to $48$ other pages. The majority of Wikipedia pages, $60\%$, are redirect pages, and $4\%$ are disambiguation pages. We estimate the total number of English articles (including disambiguation pages) to be $5.35$ million, and the total number of redirect pages to be $8.05$ million, within the confidence intervals
of the values reported by Wikipedia~\cite{WikiPage2} (Table~\ref{tab:3}).
We find the total size of English articles in Wikipedia to be $35.8$ gigabytes (GB), in reasonable agreement with the Wikipedia
statement that text alone accounts for $27.6$ GB of the storage space of English articles~\cite{WikiPage3}.
%The confidence intervals provided in Table~\ref{tab:2} can be found due to the fact that the standard errors for products of any estimators $\hat{u}$ and $\hat{v}$ can be shown to be $\sigma_{uv}^2 =\sigma_u^2\sigma_v^2 + \hat{u}^2 \sigma_v^2 + \sigma_u^2 \hat{v}^2$. %\hfill

Fig.~\ref{fig:wikipedia_fig}(a) demonstrates that the assumption of the exponential RT distribution is reasonable for Wikipedia, with some enrichment for short RTs due to the choice of network hubs as pseudotargets.
Fig.~\ref{fig:wikipedia_fig}(b) shows how the estimate of the total number of Wikipedia pages evolves as ${\cal K}_p$ increases. As in many other Internet-based networks~\cite{Faloutsos1999}, the degree distribution
of Wikipedia pages is scale-free (Fig.~\ref{fig:wikipedia_fig}(c)). In contrast, the distribution of page sizes is not scale-free, and the size of an average Wikipedia page is only $2.7$ kB (Fig.~\ref{fig:wikipedia_fig}(d), Table~\ref{tab:3}).

In conclusion, we have presented a general Bayesian approach to inferring various network properties, including its size, by using RWs that visit only a small fraction of all network nodes.
Our approach works for both weighted and unweighted undirected networks, and remains accurate in the presence of loops. Our main assumption, that of the exponentiality of the RT distribution, appears
to hold in all the cases we have examined explicitly, and can be relaxed if necessary. Our future work will focus on extending this methodology to directed and time-dependent networks.
%the estimators which arise from this theoretical framework and methodology recovered the global network properties consistently and often exactly for these five systems regardless of which of the variety of network properties we chose to determine. This suggests a very general level of applicability. Although this fidelity could certainly break down if the distribution of RTs is not well approximated by an exponential distribution. Another assumption behind this method of global network property acquisition is that the small region which the walker has explored is a representative sample of the global structure. This has clearly been the case for the systems we examined; however, this is in no way guaranteed. Even so, the model systems we chose were indeed designed to reflect real-world complex networks \cite{Erdos1961, Albert2002, Watts1998}. The next logical step in this framework is to generalize the theoretical background to include networks with directed edges allowing for the analysis of all system types in which a RW can be performed.

%\bibliography{../BayesInferNetworkStats}{}

\begin{thebibliography}{10}
\bibitem{SM2018}
See {S}upplemental {M}aterial at [{URL} will be inserted by publisher] for
  additional network inference results.
\end{thebibliography}
%\bibliographystyle{unsrt}

\end{document}